\documentstyle [12pt,epsfig,aaspp4] {article}

\newcommand{\be}{\begin{eqnarray}}
\newcommand{\ee}{\end{eqnarray}}
\newcommand{\ba}{\begin{eqnarray}}
\newcommand{\ea}{\end{eqnarray}}

\newcommand{\msun}{M_\odot}
\def\rsun{R_\odot}
\newcommand{\gsim}{\mathrel{\hbox{\rlap{\lower.55ex \hbox {$\sim$}}
                   \kern-.3em \raise.4ex \hbox{$>$}}}}
\newcommand{\lsim}{\mathrel{\hbox{\rlap{\lower.55ex \hbox {$\sim$}}
                   \kern-.3em \raise.4ex \hbox{$<$}}}}

\def\BH{{\rm BH}}
\def\SN{{\rm SN}}
\def\L{{\rm L}}

\def\D{{\rm D}}
\def\Sch{{\rm Sch}}
\lefthead{Brown, Lee, Wijers, Lee, Israelian, Bethe}
\righthead{A Theory of Gamma-Ray Bursts}

\begin{document}
\leftline{Draft Version \today}

\title{A Theory of Gamma-Ray Bursts}
\author{G. E. Brown$^1$, C.-H. Lee$^1$, R. A. M. J. Wijers$^1$,
        H. K. Lee$^2$, G. Israelian$^3$,\\ and H. A. Bethe$^4$}
\affil{
    $^1$Department of Physics \& Astronomy,
        State University of New York,\\
        Stony Brook, New York 11794, USA\\
    $^2$Department of Physics, Hanyang University, Seoul 133-791, Korea\\
    $^3$Instituto de Astrofisica de Canarias, E-38200 La Laguna,
        Tenerife, Spain\\
    $^4$Floyd R. Newman Laboratory of Nuclear Studies,
       Cornell University,\\ Ithaca, New York 14853, USA
}

\begin{abstract}
Recent observations and theoretical considerations have linked gamma-ray
 bursts with ultra-bright type Ibc supernovae (`hypernovae'). We here
work out a specific scenario for this connection. Based on earlier work,
 we argue that especially the longest bursts must be powered by the
Blandford-Znajek mechanism of electromagnetic extraction of spin energy
from a black hole. Such a mechanism requires a high angular momentum in
the progenitor object. The observed association of gamma-ray bursts with
type Ibc supernovae leads us to consider massive helium stars that form
black holes at the end of their lives as progenitors. In our analysis
we combine the numerical work of MacFadyen \& Woosley with analytic
calculations in Kerr geometry, to show that about $10^{53}$\,erg each are
available to drive the fast GRB ejecta and the supernova. The GRB ejecta
are driven by the power output through the open field lines threading
the black hole, whereas the supernova can be powered both by the shocks
driven into the envelope by the jet, and by the power delivered into the
disk via field lines connecting the disk with the black hole. We also
present a much simplified approximate derivation of these energetics.

Helium stars that leave massive black-hole remnants can only be made
in fairly specific binary evolution scenarios, namely the kind that
also leads to the formation of soft X-ray transients with black-hole
primaries, or in very massive WNL stars. Since the binary progenitors
will inevitably possess the high angular momentum we need, we propose a
natural link between black-hole transients and gamma-ray bursts. Recent
 observations of one such transient, GRO\,J1655$-$40/Nova Scorpii
1994, explicitly support this connection: its high space velocity
indicates that substantial mass was ejected in the formation of the
black hole, and the overabundance of $\alpha$-nuclei, especially
sulphur, indicates that the explosion energy was extreme, as in
SN\,1998bw/GRB\,980425. Furthermore, X-ray studies of this object
indicate that the black hole may still be spinning quite rapidly, as
expected in our model. We also show that the presence of a disk during
the powering of the GRB and the explosion is required to deposit enough
of the $\alpha$ nuclei on the companion.

\medskip

\end{abstract}

\section{Introduction}
\label{intro}

The discovery of afterglows to gamma-ray bursts has greatly increased the
possibility of studying their physics. Since these afterglows have thus
far only been seen for long gamma-ray bursts (duration $\gsim 2$\,s), we
shall concentrate on the mechanism for this subclass. The shorter bursts
(duration $\lsim 2$\,s) may have a different origin; specifically, it
has been suggested that they are the result of compact-object mergers
and therefore offer the intriguing possibility of associated outbursts
of gravity waves. (Traditionally, binary neutron stars have been 
considered in this category (Eichler et~al.\ 1989, Janka et~al.\ 1999).
More recently, Bethe \& Brown
(1998)\nocite{Bethe98} have shown that low-mass black-hole, 
neutron-star binaries, which have a ten times greater formation rate and 
are stronger gravity-wave emitters, may be the more promising source of
this kind.)

An important recent clue to the origin of long bursts is the
probable association of some of them with ultra-bright type Ibc supernovae
(Galama et~al.\ 1998, Bloom et~al.\ 1999, Galama et~al.\
2000)\nocite{Galama98C,Bloom99C,Galama00A}.  The very large explosion
energy\footnote{H\"oflich et~al.\ (1999)\nocite{Hoeflich99} have proposed
   that the explosion energy was not much larger than usual, but that the
   explosion was very asymmetric; this model also provides a reasonable fit
   to the light curve of SN\,1998bw.}
implied by fitting the light curve of SN\,1998bw, 
which was associated with
GRB\,980425\nocite{Galama98C}, indicates that a black hole was formed
in this event (Iwamoto et~al.\ 1998)\nocite{Iwamoto98}. This provides
two good pieces of astrophysical information: it implicates
black holes in the origin of gamma-ray bursts, and it demonstrates that
a massive star can explode as a supernova even if its core
collapses into a black hole.

In this paper, we start from the viewpoint that the gamma-ray burst is
powered by electromagnetic energy extraction from a spinning black hole,
the so-called Blandford-Znajek (1977)\nocite{Blandford77} mechanism.  This
was worked out in detail by Lee, Wijers, \& Brown (1999)\nocite{Lee99A},
and further details and comments were discussed by Lee, Brown, \& Wijers
(2000)\nocite{Lee99B}, who
built on work by Thorne et~al.\ (1986)\nocite{Thorne86} and Li
(2000)\nocite{Li00}.  They have shown that with the circuitry
in a 3$+$1 dimensional description using the Boyer-Lindquist metric,
one can have a simple pictorial model for the BZ mechanism.

The simple circuitry which involves steady state current flow is,
however, inadequate for describing dissipation of the black hole
rotational energy into the accretion disk formed from the original
helium envelope. In this case the more rapidly rotating black hole
tries to spin up the inner accretion disk through the closed field
lines coupling the black hole and disk. Electric and magnetic fields vary
wildly with time. Using the work of Blandford \& Spruit (2000) we
show that this dissipation occurs in an oscillatory fashion, giving
a fine structure to the GRB, and that the total dissipation should
furnish an energy comparable to that of the GRB to the accretion disk.
We use this energy to drive the hypernova explosion.

Not any black-hole system will be suitable for making GRB: the
black hole must spin rapidly enough and be embedded in a strong
magnetic field. Moreover, the formation rate must be high enough
to get the right rate of GRB even after accounting for substantial
collimation of GRB outflows. We explore a variety of models, and
give arguments why some will have sufficient energy and
extraction efficiency to power a GRB and a hypernova. We argue
that the systems known as black-hole transients are the relics of
GRBs, and discuss the recent evidence from high space velocities
and chemical abundance anomalies that these objects are relics of
hypernovae and GRBs; we especially highlight the case of
Nova Scorpii 1994 (GRO\,J1655$-$40).

The plan of this paper is as follows. We first show
that it is reasonable to expect similar energy depositions into the GRB
outflow and the accretion disk (Sect.~\ref{circuit}) and discuss the
amount of available energy to be extracted (Sect.~\ref{energy}). Then
we show the agreement of those results with the detailed numerical
simulations by MacFadyen \& Woosley, and use those simulations to
firm up our numbers (Sect.~\ref{compare}). We continue by presenting
a simple
derivation of the energetics that approximates the full results well
(Sect.~\ref{simple}).
Finally, we discuss 
some previously suggested progenitors (Sect.~\ref{progenitor})
and present our preferred progenitors: soft X-ray transients
(Sect.~\ref{sec7}).

\section{Simple Circuitry}
\label{circuit}

Although our numbers are based on the detailed review of Lee,
Wijers, \& Brown (1999)\nocite{Lee99A},
which confirms the original Blandford-Znajek
(1977)\nocite{Blandford77}
paper, we illustrate our arguments with the pictorial
treatment of Thorne et~al.\ (1986)\nocite{Thorne86} in {\it ``The Membrane
Paradigm"}. Considering the time as universal in the
Boyer-Lindquist metric, essential electromagnetic and statistical
mechanics relations apply in their 3$+$1 dimensional manifold. We
summarize their picture in our Fig.~\ref{bzcirc}.

The surface of the black hole can be considered as a conductor with surface
resistance $R_{\BH}=4\pi/c=377$ ohms. A circuit that rotates rigidly
with the black hole can be drawn from the loading region, the
low-field region up the axis of rotation of the black hole in
which the power to run the GRB is delivered, down a magnetic field
line, then from the North pole of the black hole along the
(stretched) horizon to its equator. From the equator we continue
the circuit through part of the disk and then connect it upwards
with the loading region. We can also draw circuits starting from
the loading region which pass along only the black hole or go
through only the disk, but adding these would not change the
results of our schematic model.

Using Faraday's law, the voltage $V$ can be found by integrating
the vector product of charge velocity, $\vec v$, and magnetic
field, $\vec B$, along the circuit:
    \be
    V =\int
    [\vec v\times\vec B]\cdot d{\vec l},
    \ee
($d{\vec l}$ is the line element along the circuit). Because this law
involves $\vec v\times \vec B$ the integrals along the field lines
make no contribution. We do get a contribution $V$
from the integral from North
pole to equator along the black hole surface. Further
contributions to $V$ will come from cutting the field lines from
the disk. We assume the field to be weak enough in the loading
region to be neglected.

The GRB power, $E_{\rm GRB}$, will be
   \be
   \dot E_{\rm GRB}=I_{\BH+\D}^2 R_\L
   \ee
where $R_\L$ is the resistance of the loading region, and the current
is given by
   \be
   I_{\BH+\D}^2 =\left(\frac{V_\D+V_{\BH}}{R_\D+R_{\BH}+R_\L}\right)^2.
   \ee
(The index BH refers to the black hole, L to the load region, and D to
the disk.)

The load resistance has been estimated in various ways and for various
assumptions by Lovelace, MacAuslan, \& Burns (1979)\nocite{Lovelace79}
and by MacDonald \& Thorne (1982)\nocite{MacDonald82}, and by Phinney
(1983)\nocite{Phinney83}. All estimates agree that to within a factor
of order unity $R_\L$ is equal to $R_\BH$.

In a similar fashion, some power will be deposited into the disk
    \be
    \dot E_{disk}=I_{BH+D}^2 R_D
    \ee
but this equilibrium contribution will be small because of the low
disk resistance $R_D$.

Blandford \& Spruit (2000) have shown that important dissipation into
the disk comes through magnetic field lines coupling the disk to the
black hole rotation. As shown in Fig.~\ref{mbreak} these lines,
anchored in the inner disk, thread the black hole.

The more rapidly rotating black hole will provide torques, along its
rotation axis, which spin up the inner accretion disk, in which the closed
magnetic field lines are anchored. With increasing centrifugal force
the material in the inner disk will move outwards, cutting down the
accretion. Angular momentum is then advected outwards, so that the
matter can drift back inwards. It then delivers more matter to
the black hole and is flung outwards again. The situation is like that of
a ball in a roulette wheel (R.D. Blandford, private communication).
First of all it is flung outwards and then drifts slowly inwards.
When it hits the hub it is again thrown outwards.
The viscous inflow time for the fluctuations is easily estimated to be
   \be
   \tau_d\sim \Omega^{-1}_{disk}
   \left(\frac{r}{H}\right)^2 \alpha_{vis}^{-1}
   \ee
where $H$ is the height of the disk at radius $r$, $\Omega_{disk}$ its
angular velocity, and $\alpha_{vis}$
is the usual $\alpha$-parameterization of the viscosity. We choose
$\alpha_{vis}\sim 0.1$, $r/H\sim 10$ for a thin disk and then arrive at
$\tau_d\sim 0.1$\,s. We therefore expect variability on all time scales
between the Kepler time (sub-millisecond) and the viscous time, which may
explain the very erratic light curves of many GRBs.

We suggest that the GRB can be powered by $\dot E_{\rm GRB}$ and a
Type Ibc supernova explosion by $\dot E_\SN$ where $\dot E_{\SN}$
is the power delivered through dissipation into the disk. To the
extent that the number of closed field lines coupling disk and
black hole is equal to the number of open field lines threading
the latter, the two energies will be equal. In the spectacular
case of GRB 980326 (Bloom et al. 1999)\nocite{Bloom99C}, the GRB
lasts about $5$\,s, which we take to be the time that the central
engine operates. We shall show that up to $\sim 10^{53}$\,erg is
available to be delivered into the GRB and into the accretion
disk, the latter helping to power the supernova (SN) explosion.
This is more energy than needed and we suggest that
injection of energy into the disk shuts off the central engine by
blowing up the disk and thus removing the magnetic field needed for the
energy extraction from the black hole. If
the magnetic field is high enough the energy will be delivered in
a short time, and the quick removal of the disk will leave the black hole
still spinning quite rapidly.

\section{Energetics of GRBs}
\label{energy}

The maximum energy that can be extracted from the BZ mechanism (Lee, Wijers,
\& Brown 1999)\nocite{Lee99A} is
   \be
   (E_{BZ})_{max} \simeq 0.09\ M_\BH c^2.
   \ee
This is 31\% of the black hole rotational energy, the remainder going
toward increasing the entropy of the black hole. This maximum energy
is obtained if the extraction efficiency is
   \be
   \epsilon_\Omega
   =\frac{\Omega_{disk}}{\Omega_{H}}=0.5.
   \label{eqeps}
   \ee
In Appendix A we give numerical estimates
for this ratio for various $\omega=\Omega_{disk}/\Omega_{K}$ and various
radii in the region of parameter space we consider. As explained in
Section \ref{circuit} we expect the material in the inner disk to
swing in and out around the marginally stable radius,
$r_{ms}$. It can be seen from
the Table \ref{appBtab1} and Appendix A
that the relevant values of $\epsilon_\Omega$
are close to that of eq.~(\ref{eqeps}).

For a $7\msun$ black hole, such as that found in Nova Sco 1994
(GRO\,J1655$-$40),
   \be
   E_{max} \simeq 1.1\times 10^{54}\ {\rm erg}.
   \label{eq5}\ee
We estimate below that the energy available in a typical case will
be an order of magnitude less than this.
Without collimation, the estimated gamma-ray energy in GRB\,990123 is
about $4.5\times 10^{54}$\,erg
(Andersen et~al.\ 1999)\nocite{Andersen99}.
The BZ scenario entails substantial beaming, so this energy should be multiplied
by $d\Omega/4\pi$, which may be a small factor (perhaps $10^{-2}$).

The BZ power can be delivered at a maximum rate of
    \be
    P_{\rm BZ} =6.7\times 10^{50}\left(\frac{B}{10^{15}{\rm G}}\right)^2
    \left(\frac{M_\BH}{\msun}\right)^2\ {\rm erg\ s^{-1}},
    \label{eq7}
    \ee
(Lee et~al.\ 1999)\nocite{Lee99A}
so that high magnetic fields are necessary for rapid delivery.

The above concerns the maximum energy output into the jet and the disk.
The real energy available in black-hole spin in any given case, and the
efficiency with which it can be extracted, depend on the rotation
frequency of the newly formed black hole and the disk or torus around it.
The state of the accretion disk around the newly formed black hole, and
the angular momentum of the black hole, are somewhat uncertain. However,
the conditions should be bracketed between a purely Keplerian, thin disk
(if neutrino cooling is efficient) and a thick, non-cooling hypercritical
advection-dominated accretion disk (HADAF), of which we have a model
(Brown, Lee \& Bethe 2000)\nocite{Brown00}. Let us examine the result
for the Keplerian case.
In terms of
   \be
   \tilde a\equiv\frac{Jc}{M^2 G},
   \ee
where $J$ is the angular momentum of the black hole,
we find the rotational energy of a black hole to be
   \be \label{erotfull}
   E_{rot} = f(\tilde a) M c^2,
   \ee
where
   \be
   f(\tilde a)=1-\sqrt{\frac 12 (1+\sqrt{1-{\tilde a}^2})}.
   \label{eqfa}
   \ee
For a maximally rotating black hole one has $\tilde{a}=1$\footnote{
   As an aside, we note a nice mnemonic: if we define a velocity $v$
   from the black-hole angular momentum by $J=MR_\Sch v$, so that $v$
   carries the quasi-interpretation of a rotation velocity at the
   horizon, then $\tilde{a}=2v/c$. A maximal Kerr hole, which has
   $R_{\rm event}=R_\Sch/2$, thus has $v=c$. For $\tilde{a}\lsim0.5$,
   the rotation energy is well approximated by the easy-to-remember
   expression $E_{\rm rot} = \frac{1}{2}Mv^2$.}.

We begin with a neutron star in the middle of a Keplerian
accretion disk, and let it 
accrete enough matter to send it into a black hole.
In matter free regions the last stable orbit of a particle around
a black hole in Schwarzschild geometry is
    \be
    r_{\rm lso}=3 R_{\Sch} =6\frac{GM}{c^2}.
    \ee
This is the marginally stable orbit $r_{\rm ms}$.
However, under conditions of hypercritical accretion,
the pressure and energy profiles are changed and it is better
to use (Abramowicz et al. 1988)\nocite{Abramowicz88}
    \be
    r_{\rm lso}\gsim 2 R_{\Sch}.
    \ee
With the equal sign we have the marginally bound orbit $r_{\rm mb}$.
With high rates of accretion we expect this to be a  good approximation
to $r_{\rm lso}$.
The accretion disk can be taken to extend down to the last stable
orbit (refer to Appendix B for the details).

We take the angular velocity to be Keplerian, so that the disk velocity
$v$ at radius $2 R_{\Sch}$ is given by
    \be
    v^2 = \frac{GM}{2 R_{\Sch}}=\frac{c^2}{4},
    \ee
or $v=c/2$. The specific angular momentum, $l$, is then
    \be
    l \ge 2 R_{\Sch} v = 2\frac{GM}{c}
    \label{eq16}\ee
which in Kerr geometry indicates $\tilde a\sim 1$. Had we taken one
of the slowest-rotating disk flows that are possible, the advection-dominated
or HADAF case (Narayan and Yi 1994, 
Brown, Lee \& Bethe 2000\nocite{Brown00}), which
has $\Omega^2=2\Omega_K^2/7$, we would have arrived at $\tilde a\sim 0.54$,
so the Kerr parameter will always be high.

Further accretion will add angular momentum to the black hole at a rate
determined by the angular velocity of the inner disk.
The material accreting into the
black hole is released by the disk at $r_{\rm lso}$, where the angular
momentum delivered to the black hole is determined. This angular
momentum is, however, delivered into the black hole at the event
horizon $R_{\Sch}$, with velocity at least double that at which it
is released by the disk, since the lever arm at the event horizon
is only half of that at $R_{\Sch}$, and angular momentum is
conserved. With more rapid
rotation involving movement towards a Kerr geometry where the
event horizon and last stable orbit coincide at
   \be
   r_{\rm lso}=R_{\rm event}=\frac{GM}{c^2}.
   \ee
Although we must switch over to a Kerr geometry for quantitative results,
we see that $\tilde a$ will not be far from its maximum value of unity.
Again, for the lower angular-momentum case of a HADAF, the expected
black-hole spin is not much less.

\section{Comparison with Numerical Calculation}
\label{compare}

Our schematic model has the advantage over numerical calculations
that one can see analytically how the scenario changes with change
in parameters or assumptions. However, our model is useful only if
it reproduces faithfully the results of more complete calculations
which involve other effects and much more detail than we include.
We here make comparison with Fig.19 of MacFadyen \& Woosley
(1999)\nocite{MacFadyen99}.
Accretion rates, etc., can be read off from their figure
which we reproduce as our Fig.\ref{woosley}. MacFadyen \& Woosley
prefer $\tilde{a}_{initial}=0.5$ (We have removed their curve for
$\tilde{a}_{initial}=0$). This is a reasonable value if the black hole
forms from a contracting proto-neutron star near breakup.
MacFadyen \& Woosley find that $\tilde{a}_{initial}=0.5$ is more
consistent with the angular momentum assumed for the mantle than
$\tilde{a}_{initial}=0$. (They take the initial black hole to have mass
$2\msun$; we choose the Brown \& Bethe (1994)\nocite{Brown94}
mass of $1.5\msun$.)
We confirm this in the next section.

After 5 seconds (the duration of GRB\,980326) the MacFadyen \&
Woosley black hole mass is $\sim 3.2\msun$ and their Kerr
parameter $\tilde{a}\sim 0.8$, which gives $f(\tilde a)$ of our
eq.(\ref{eqfa}) of 0.11. With these parameters
we find $E=2\times 10^{53}$ erg,
available for the GRB and SN explosion.

One can imagine that continuation of the MacFadyen \& Woosley
curve for $M_{BH}(\msun)$ would ultimately give something like our
$\sim 7\msun$, but the final black hole mass may not be relevant
for our considerations. This is because more than enough energy is
available to power the supernova in the first 5 seconds; as the
disk is disrupted, the magnetic fields supported by it will also
disappear, which turns off the Blandford-Znajek mechanism.

Power is delivered at the rate given by eq.(\ref{eq7}). Taking a 
black hole mass relevant here, $\sim 3.2\msun$, we require a 
field strength of $\sim 5.8\times 10^{15}$\,G 
in order for our estimated energy ($4\times 10^{52}$ erg) to be
delivered in 5\,s (the duration of GRB\,980326).
For such a relatively short burst, we see that the required field 
is quite large, but it is still not excessive if we bear in mind that
magnetic fields of $\sim10^{15}$\,G have already been observed in
magnetars (Kouveliotou 1998, 1999)\nocite{Kouveliotou98,Kouveliotou99A}.
Since in our scenario we have many more progenitors than 
there are GRBs, we suggest that the necessary fields are obtained only in
a fraction of all potential progenitors.

Thus we have an extremely simple scenario for powering a GRB and the
concomitant SN explosion in the black hole transients, which we will
discuss in Section.\ref{sec7.2}. After the first second the newly
evolved black hole has $\sim 10^{53}$ erg of rotational energy
available to power these. The time scale for delivery of this
energy depends (inversely quadratically) on the magnitude of the
magnetic field in the neighborhood of the black hole, essentially
that on the inner accretion disk. The developing supernova
explosion disrupts the accretion disk; this removes the magnetic fields
anchored in the disk, and self-limits the energy the B-Z mechanism can deliver.

\section{An Even More Schematic Model}
\label{simple}

Here we calculate the energy available in a rotating black
hole just after its birth (before accretion
adds more). Our model is to take a $1.5\msun$ neutron star which
co-rotates with the inner edge of the accretion disk in which it is
embedded. The neutron star then collapses to a black hole, conserving
its angular momentum. Since the accretion disk is neutrino cooled,
but perhaps not fully thin, its angular velocity will be somewhere
between the HADAF value and the Keplerian value. We parameterize it
as $\Omega=\omega\Omega_{\rm K}$, where $\omega=1$ for Keplerian and
$\omega=\sqrt{2/7}\sim0.53$ for the HADAF.

The moment of inertia, $I$,
of a neutron star is well fitted for many different
equations of state with the simple expression
\be
  I = \frac{0.21\,MR^2}{1-2GM/Rc^2}
\ee
(Lattimer \& Prakash 2000)\nocite{Lattimer00}. With $J=\omega I\Omega_{\rm K}$
and a neutron star of $1.5\msun$, with a radius of 10\,km, we find
\be
  \tilde{a}^2 = \left(\frac{Jc}{GM^2}\right)^2 = 0.64\omega^2.
\ee
We choose $\omega \simeq 1.0$ to roughly reproduce the MacFadyen \& Woosley
value of $\tilde{a}$, see our Fig.~\ref{woosley}.
We do not really believe the disk to be so efficiently neutrino cooled
that its angular velocity is Keplerian; i.e. $\omega=1$, but it may be not
far from it. Our $\omega$ should be more properly viewed as a fudge
factor which allows us to match the more complete MacFadyen \& Woosley
calculation.
MacFadyen \& Woosley find that, while
the accretion disk onto the black hole is forming, an additional solar mass
of material is added to it ``as the dense stellar core collapses
through the inner boundary at all polar angles''. We shall add this to
our $1.5\msun$ and take the black hole mass to be $2.5\msun$. We neglect the
increase in spin of the black hole by the newly accreted matter; this is
already included in the MacFadyen \& Woosley results. For
$\tilde{a}^2 = 0.64$ we find $f(\tilde{a}^2)=0.11$, so that the
black hole rotation energy becomes
\be
  E_{\rm BZ}= 1.5\times 10^{53} \ {\rm erg}
\ee
in rough agreement with the estimates of MacFadyen \& Woosley in the
last section.



%
\section{Previous Models}
\label{progenitor}

      \subsection{Collapsar}

We have not discussed the Collapsar model of Woosley
(1993)\nocite{Woosley93B}, and MacFadyen \& Woosley
(1999)\nocite{MacFadyen99}. In this model the center of a rotating
Wolf-Rayet star evolves into a black hole, the outer part being held
out by centrifugal force. The latter evolves into an accretion disk and
then by hypercritical accretion spins the black hole up. MacFadyen \&
Woosley point out that ``If the helium core is braked by a magnetic field
prior to the supernova explosion to the extent described by Spruit \&
Phinney (1998)\nocite{Spruit98} then our model will not work for single
stars." Spruit \& Phinney argue that magnetic fields maintained by
differential rotation between the core and envelope of the star will
keep the whole star in a state of approximately uniform rotation until
10 years before its collapse.  As noted in the last section, with the
extremely high magnetic fields we need the viscosity would be expected
to be exceptionally high, making the Spruit \& Phinney scenario probable.
Livio \& Pringle (1998)\nocite{Livio98} have commented that one finds
evidence in novae that the coupling between layers of the star by magnetic
fields may be greatly suppressed relative to what Spruit \& Phinney assumed.
However, we note that  even with this suppressed coupling, they find
pulsar periods from core collapse supernovae no shorter than 0.1\,s.
Independent evidence for the fact that stellar cores mostly rotate no faster
than this comes from the study of supernova remnants: Bhattacharya (1990, 1991)
concludes that the absence of bright, pulsar-powered plerions in most SNRs
indicates that typically pulsar spin periods at birth are no shorter than
0.03--0.05\,s.  Translated to our black holes, such spin periods would imply
$\tilde{a}\lsim0.01$, quite insufficient to power a GRB.
As a cautionary note, we might add that without magnetic coupling the cores
of evolved stars can spin quite rapidly (Heger et~al.\  
2000)\nocite{Heger00}. This rapid initial spin may be reconciled with
Bhattacharya's limit if r-mode instabilities cause very rapid spindown
in the first few years of the life of a neutron star (e.g., Heger, Langer,
\& Woosley 2000, Lindblom \& Owen 1999)\nocite{Lindblom99}.

      \subsection{Coalescing Low-Mass Black Holes and Helium Stars}
      \label{coalesc}

Fryer \& Woosley (1998)\nocite{Fryer98}
suggested the scenario of a black hole spiraling
into a helium star. This is an efficient way to spin up the
black hole.

Bethe \& Brown (1998)\nocite{Bethe98}
evolved low-mass black holes with helium star
companion, as well as binaries of compact objects. In a total available
range of binary separation $0.04 < a_{13}<4$, low-mass black-hole,
neutron-star binaries were formed when $0.5<a_{13}<1.4$ where
$a_{13}$ is the initial binary separation in units of $10^{13}$ cm.
The low-mass black hole coalesces with the helium star in the range
$0.04<a_{13}<0.5$. Binaries were distributed logarithmically in $a$.
Thus, coalescences are more common than low-mass black-hole,
neutron-star binaries by a factor of $\ln(0.5/0.04)/\ln(1.9/0.5)=1.9$

In Bethe \& Brown (1998), the He-star, compact-object binary was
disrupted $\sim 50\%$ of the time by the He-star explosion.
This does not apply to the coalescence. Thus, the rate of
low-mass black-hole, He-star mergers is 3.8 times the formation
rate of low-mass black-hole, neutron-star binaries, or
   \be
   R=3.8\times 10^{-4}\ {\rm yr}^{-1}
   \ee
in the Galaxy. The estimated empirical rate of GRBs, with a factor
of 100 for beaming, is $10^{-5}$ yr$^{-1}$ in the Galaxy
(Appendix C of Brown et~al.\ 1999)\nocite{Brown99B}. Thus, the
number of progenitors is more than adequate.

In Bethe \& Brown (1998)\nocite{Bethe98}
the typical black hole mass was $\sim 2.4
M_\odot$, somewhat more massive than their maximum assumed neutron
star mass of $1.5 M_\odot$. As it enters the helium star companion
an accretion disk is soon set up and the accretion scenario will
follow that described above, with rotating black holes of various
masses formed. Brown, Lee, \& Bethe (2000) find that the black hole will 
be spun up quickly. We have not pursued this scenario beyond the point that
it was developed by Fryer \& Woosley (1998).

\section{Soft X-ray Transients as Relics of Hypernovae and GRB}
 \label{sec7}

\subsection{Our Model: Angular Momentum}
\label{sec7.1}

We favor a model of hypernovae similar to MacFadyen \& Woosley
(1999) in that it involves a failed supernova as a centerpiece.
But, in distinction to MacFadyen \& Woosley, our initial system is
a binary, consisting of a massive star A (which will later become
the failed SN) and a lighter companion B, which serves to provide 
ample angular momentum.

Failed supernovae require a ZAMS mass of $20-35\msun$, according
to the calculations of Woosley \& Weaver (1995) as interpreted by
Brown, Lee, \& Bethe (1999). The limits 20 and $35\msun$ are not
accurately known, but it is a fairly narrow range, so we shall in
many of our calculations assume a ``typical" ZAMS mass of
$25\msun$. The heavy star A must not be in a {\it close} binary
because then its hydrogen envelope would be removed early in its
evolution and therefore the star would lose mass by wind at a very
early stage and become a low-mass compact object (Brown,
Weingartner, \& Wijers 1996). Instead, we assume a wide binary, with
a separation, $a$ in the range
  \be
  a=500-1000\rsun,
    \label{eq7.2}
  \ee
so star A evolves essentially as a single star through its first
few burning stages. It is essential
that most of the He core burning is completed before its hydrogen
envelope is removed (Wellstein \& Langer 1999; Heger \& Wellstein
2000). We assume the initial distance $a$ between the
two stars to be in this range. When star A fills its Roche lobe, the
companion, star B, will spiral inwards.

The initiation and early development of the common envelope has been
best treated by Rasio \& Livio (1996)\nocite{Rasio96}.
This is the only phase that can at
present be modeled in a realistic way. They find a short viscous time in
the envelope, but emphasize that numerical viscosity may play an important
role in their results. However, we believe the viscosity to be large.
Torkelsson et al. (1996)\nocite{Torkelsson96}
showed  the Shakura-Sunyaev (1973)\nocite{Shakura73} viscosity parameter,
$\alpha_{SS}$,
to range from 0.001 to 0.7, with the higher values following from the
presence of vertical magnetic fields. Since in our Blandford-Znajek
model extremely high magnetic fields $\sim 10^{15}$ G are needed
in the He envelope to deliver the energy rapidly, we believe $\alpha_{SS}$
to be not much less than unity.
Given such high viscosities, it seems reasonable to follow the
Rasio-Livio extrapolation, based on a short viscous transport time,
to later times.
The most significant new result of Rasio \& Livio ``is that,
during the dynamical phase of common envelope evolution, a corotating
region of gas is established near the central binary.
The corotating region has the shape of an oblate spheroid encasing the
binary (i.e., the corotating gas is concentrated in the orbital plane).''

A helium core, which we deal with, is not included in their calculations,
because they do not resolve the inner part of the star numerically.
However, since the physics of the spiral-in does not really change as it
proceeds past the end of their calculations, it seems most likely that
during further spiral-in, the spin-up of material inside the orbit of the
companion will continue to be significant.

Star B will stop spiraling in when it has ejected the H envelope
of A. Since we assume that all stars A have about the same mass,
and that $a_i$ is very large, we expect
   \be
   \frac{M_B}{a_f}\simeq {\rm const.}
    \label{eq7.3}
   \ee
From section~\ref{sec7.2} we conclude that $a_f$ is a few $\rsun$
for $M_B=(0.4-1)\msun$. Now the He cores of stars of ZAMS mass
$M=20-35\msun$ have a radius about equal to $\rsun$. Therefore
small $M_B$ stars will spiral into the He core of A. There they
cannot be stopped but will coalesce with star A. However, they
will have transmitted their angular momentum to star A.

Star B of larger mass will stop at larger $a_f\gg\rsun$. It is
then not clear whether they will transfer all of their angular
momentum to star A. In any case, they must generally wait until
they evolve off the main sequence into the subgiant or possibly even
the giant stage before they can fill their Roche Lobes and later
accrete onto the black hole resulting from star A.

The Kepler velocity of star B at $a_f$ is
   \be
   V_K^2=G \frac{M_{a_f}}{a_f}.
   \ee
We estimate the final mass of A, after removal of its hydrogen
envelope, to be about $10\msun$; then
  \be
  V_K\simeq 1.2\times 10^8 a_{f,11}^{-1/2}{\,\rm cm\,s}^{-1},
  \label{eq7.5}
  \ee
where $a_{f,11}$ is $a_f$ in units of $10^{11}$ cm. The specific
angular momentum of B is then
   \be
   j(B)= a_f V_K=1.2\times 10^{19} a_{f,11}^{1/2}{\,\rm cm^2 \ s^{-1}}.
   \label{eq7.6}
   \ee
If B and A share their angular momentum, the specific angular
momentum is reduced by a factor $M_B/(M_{A,f}+M_B)$ which we
estimate to be $\sim 0.1$. Since $a_f$ should be $\gsim 3 \rsun$ (See
Table~\ref{tabhans}), the specific angular momentum of A should be
   \be
   j(A)\gsim 10^{18} {\,\rm cm^2 \ s^{-1}}.
   \label{eq7.7}
   \ee
Star B has now done its job and can be disregarded.

\subsection{Supernova and collapse}
\label{sec7.2}

Star A now goes through its normal evolution, ending up as a
supernova. But since we have chosen its mass to be between 20 and
$35\msun$, the SN shock cannot penetrate the heavy envelope but is
stopped at some radius
  \be
  R_{\SN}\simeq 10^{10} {\rm cm},
    \label{eq7.8}
  \ee
well inside the outer edge of the He envelope. We estimate $ R_{\SN}$
by scaling from SN\,1987A: in that supernova, with progenitor mass
$\sim18\,\msun$, most of the He envelope was returned to the galaxy. The
separation between compact object and ejecta was estimated to occur at
$R\sim5\times10^8$\,cm (Woosley 1988, Bethe 1990) at mass point 1.5\,$\msun$
(gravitational). Woosley and Weaver (1995) find remnant masses of
$\sim2\,\msun$, although with large fluctuations, for ZAMS masses in the
range 20--35$\msun$, which go into high-mass black holes. From table 3 of
Brown, Weingartner, and Wijers (1996) we see that fallback between 
$R=3.5$ and $4.5\times10^8$\,cm is 0.03$\msun$. Using this we can extrapolate
to $R=10^{10}$\,cm as the distance within which matter has to begin falling in
immediately in our heavier stars, to make up a compact object of 2\,$\msun$.
Unlike in 1987A the shock energy in the more massive star does not suffice
to eject the envelope beyond this point, and the remaining outer envelope
will also eventually fall back.

At $R_{\SN}$, the specific angular momentum of Kepler motion around
a central star of mass $10\msun$ is, cf. eq.(\ref{eq7.6})
   \be
   j_K(10\msun)=1.2\times 10^{19} R_{f,11}^{1/2}{\,\rm cm^2\ s^{-1}}
   =4\times 10^{18}{\,\rm cm^2\ s^{-1}}.
   \label{eq7.9}
   \ee
In reality, at this time the central object has a mass $M\sim
1.5\msun$ (being a neutron star) and since $j_K\sim V_K\sim
M^{1/2}$
   \be
   j_K(1.5\msun) =1.5\times 10^{18} {\rm cm^2\  s^{-1}}.
    \label{eq7.10}
   \ee
The angular momentum inherent in star A, eq.(\ref{eq7.7}), is
therefore greater than the Kepler angular momentum. This would not
be the case had our initial object been a single star, a
collapsar. (The collapsar may work none the less, but our binary
model is more certain to work.)

The supernova material is supported by pressure inside the cavity,
probably mostly due to electromagnetic radiation. The cavity inside
$R_\SN$ is rather free of matter. After a while, the pressure in
the cavity will reduce. This may happen by opening toward the
poles, in which case the outflowing pressure will drive out the
matter near the poles and create the vacuum required for the gamma
ray burst. Reduction of pressure will also happen by neutrino
emission.
As the pressure gets reduced, the SN material will fall in toward
the neutron star in the center. But because the angular momentum
of the SN material is large (eq.\ref{eq7.7}) the material must
move more or less in Kepler orbits; i.e., it must spiral in. This
is an essential point in the theory.

If $j(A)$ is less than $j_K$ at $R_\SN$, the initial motion will
have a substantial radial component in addition to the tangential one.
But as the Kepler one decreases, cf.\ eq.\ref{eq7.9}, there will
come a point of $r$ at which $j_K=j(A)$. At this point an
accretion disk will form, consisting of SN material spiraling in
toward the neutron star. The primary motion is circular, but
viscosity will provide a radial component inward
   \be
   v_r\sim \alpha v_K
   \ee
where $\alpha$ is the viscosity parameter. It has been argued by
Brandenburg et al. (1996) that $\alpha\sim 0.1$ in the presence of
equipartition magnetic fields perpendicular to the disk, and it
may be even larger with the high magnetic fields required for
GRBs. Narayan \& Yi (1994) have given analytical solutions for
such accretion disks.
The material will arrive at the neutron star essentially
tangentially, and therefore its high angular momentum will
spin up the neutron star substantially.
Accretion will soon make the neutron star collapse into a black
hole. The angular momentum will be conserved, so the angular velocity
is increased since the black hole has smaller radius than the
neutron star. Thus the black hole is born with considerable spin.

A large fraction of the material of the failed supernova will accrete onto the
black hole, giving it a mass of order $7\msun$. All this material
adds to the angular momentum of the black hole since all of it has
the Kepler velocity at the black hole radius. Our estimates show
that the black hole would be close to an extreme Kerr hole
(Section~\ref{simple}), 
were it to accrete all of this
material. It may, however, be so energetic that it drives off part
of the envelope in the explosion before it can all accrete
(see Section~\ref{simple}).

\subsection{Soft X-ray Transients with Black-Hole Primaries}

Nine binaries have been observed which are black-hole X-ray transients.
All contain a high-mass black hole, of mass $\sim 7\msun$. In
seven cases the lower-mass companion (star B) has a mass $\lsim \msun$.
The two stars are close together, their distance being 
of order $5\rsun$. Star B
fills its Roche Lobe, so it spills over some material onto the black
hole. The accretion disk near the black hole emits soft X rays. Two of
the companions are subgiants, filling their Roche lobes at a few times
larger separations from the black hole.

In fact, however, the accretion onto the central object is not constant,
so there is usually no X-ray emission. Instead, the material forms
an accretion disk around the black hole, and only when enough
material has been assembled, it falls onto the black hole to give
observable X rays. Hence, the X-ray source is transient. Recent
observation of a large space velocity of Cygnus X-1 (Nelemans et
al. 1999) suggests that it has evolved similarly to the transient
sources, with the difference that the companion to the black hole
is an $\sim 18\msun$ O star. The latter pours enough matter onto the
accretion disk so that Cyg X-1 shines continuously. We plan to
describe the evolution of Cyg X-1 in a future paper (Brown et al.
2000).

Table~\ref{tabhans} is an abbreviated list of data on transient
sources. A more complete table is given in Brown et~al.\
(1999b). Two of the steady X-ray sources, in the LMC, 
have been omitted,
because we believe the LMC to be somewhat special because of its
low metallicity; also masses, etc.,  of these two are not as well
measured. Of the others, 6 are main-sequence K stars, one is
main-sequence  M, and the other
two have masses greater than the Sun. The masses given are
geometric means of the maximum and minimum masses given by the
observers. The distance $a$ between the black hole and the optical
(visible) star is greater for the heavier stars than for the K-
and M stars (except the more evolved one of them) as was expected in
Section~\ref{sec7.1} for the spiraling in of star B. The table
also gives the radius of the Roche Lobe and the specific orbital
angular momentum of star B.

Five K stars have almost identical distance $a\sim 5\rsun$, and
also Roche Lobe sizes, $\sim 1.0\rsun$. These Roche Lobes can be
filled by K stars on the main sequence. The same is true for the
M star. Together, K and M stars cover the mass range from 0.3 to
1$\msun$. The two heavier stars have Roche Lobes of 3 and $5\rsun$
which cannot possibly be filled by main-sequence stars of mass
$\sim 2\msun$. We must therefore assume that these stars are
subgiants, in the Herzsprung gap.
These stars spend only about 1\% of their life as subgiants, so we
must expect that there are many ``silent" binaries in which the
$2\msun$ companion has not yet evolved off the main sequence and
sits well within its Roche lobe, roughly 100 times more.
The time as subgiants is even
shorter for more massive stars; this explains their absence among
the transient sources.

Therefore we expect a large  number of ``silent partners": stars
of more than $1\msun$, still on their main sequence, which are far
from filling their Roche Lobe and therefore do not transfer mass
to their black hole partners. In fact, we do not see any reason
why the companion of the black hole could not have any mass, up to
the ZAMS mass of the progenitor of the black hole; it must only
evolve following the formation of the black hole. It then crosses
the Herzsprung gap in such a short time, less than the thermal
time scale, that star A cannot accept the mass from the
companion, so that common envelope evolution must ensue. If we include
these `silent partners' in the birth rate, assuming a flat mass ratio
distribution, we enhance the total birth rate of black-hole binaries
by a factor 25 over the calculations by Brown, Lee, \& Bethe (1999).

On the lower mass end of the companions, there is only one M star.
This is explained in terms of the model of Section~\ref{sec7.1} by
the fact that stars of low mass will generally spiral into the He core of
star A, and will coalesce with A, see below eq.(\ref{eq7.3}), so
no relic is left. (Since the core is left spinning rapidly, these
complete merger cases could also be suitable GRB progenitors.)
As the outcome of the spiral-in depends also on
other factors, such as the initial orbital separation and the primary
mass, one may still have an occasional survival of an M star binary
(note that the one M star companion is M0, very nearly in
the K star range).  

The appearance of the black hole transient X-ray binaries is much
like our expectation of the relic of the binary which has made a
hypernova: a black hole of substantial mass, and an ordinary star,
possibly somewhat evolved, of smaller mass. We expect 
that star B would stop at a distance $a_f$
from star A which is greater if the mass of B is greater (see
Section~\ref{sec7.1}). This is
just what we see in the black-hole binaries: the more massive
companion stars ($\sim 2\msun$)
are further from the black hole than the K stars.
We also note that the estimated birth rate of these binaries is high
enough for them to be the progenitors of GRB, even if only in
a modest fraction of them the conditions for GRB powering are achieved.

\subsection{Nova Scorpii 1994 (GRO J1655-40)}

Nova Sco 1994 is a black hole transient X-ray source. It consists
of a black hole of $\sim 7\msun$ and a subgiant of about $2\msun$.
Their separation is $17\rsun$.
Israelian et al. (1999) have analyzed the spectrum of the subgiant
and have found that the $\alpha$-particle nuclei O, Mg, Si and S
have abundances 6 to 10 times the solar value. This indicates that
the subgiant has been enriched by the ejecta from a supernova
explosion; specifically, that some of the ejecta of the supernova
which preceded the present Nova Sco (a long time ago) were
intercepted by star B, the present subgiant. Israelian et al.\ (1999)
estimate an age since accretion started from the assumption that
enrichment has only affected the outer layers of the star. We here
reconsider this: the time that passed since the explosion of the 
progenitor of the black hole is roughly the main-sequence lifetime
of the present subgiant companion, which given its mass of $\sim$2$\msun$
will be about 1\,Gyr. This is so much longer than any plausible mixing
time in the companion that the captured supernova ejecta must by now be 
uniformly mixed into the bulk of the companion. This rather increases
the amount of ejecta that we require the companion to have captured.
(Note that the accretion rate in this binary is rather less than expected
from a subgiant donor, though the orbital period leaves no doubt that
the donor is more extended than a main-sequence star (Reg\H{o}s, Tout,
and Wickramasinghe 1998). It is conceivable that the high metal abundance
has resulted in a highly non-standard evolution of this star, in which case
one might have to reconsider its age.)

The presence of large amounts of S is particularly significant.
Nomoto et al.\ (2000) have calculated the composition of a
hypernova from an $11\msun$ CO core, see Fig.~\ref{fig1i}. This
shows substantial abundance of S in the ejecta. Ordinary supernovae
produce little of this element, as shown by the results of Nomoto
et al.\ (2000) in Fig.~\ref{fig1i}.
The large amount of S, as well as O, Mg and Si we
consider the strongest argument for considering Nova Sco 1994 as a
relic of a hypernova, and for our model, generally.

Fig.~\ref{fig1i} also shows that $^{56}$Ni and $^{52}$Fe are
confined to the inner part of the hypernova, and if the cut
between black hole and ejecta is at about $5\msun$, there will be
no Fe-type elements in the ejecta, as observed in Nova Scorpii
1994. By contrast hypernova 1998bw shows a large amount of Ni,
indicating that in this case the cut was at a lower included mass.

The massive star A in Nova Sco will have gone through a hypernova
explosion when the F-star B was still on the main sequence, its
radius about $1.5\rsun$. Since the explosion caused an expansion
of the orbit, the orbital separation $a$ was smaller at the time
of the supernova than it is now, roughly by a factor 
\be
  a_{\rm then} = a_{\rm now}/(1+\Delta M/M_{\rm now}).
\ee
($\Delta M$ is the mass lost in the explosion;
see, e.g., Verbunt, Wijers, and Burm 1990). With $\Delta M\sim0.8M_{\rm now}$,
as required by the high space velocity, this means $a_{\rm then}=10\rsun$.
Therefore the fraction of solid angle
subtended by the companion at the time of explosion was 
  \be
  \frac{\Omega}{4\pi} =\frac{\pi (1.5\rsun)^2}{4\pi (10\rsun)^2}
  \approx 6\times 10^{-3}.
    \label{eq7.11}
  \ee
Assuming the ejecta of the hypernova to have been at least
$5\msun$ (Nelemans et al. 1999), the amount deposited on star B
was
   \be
   M_D \gsim 0.03\msun.
    \label{eq7.12}
   \ee
The solar abundance of oxygen is about 0.01 by mass, so with the abundance
in the F star being 10 times solar, and oxygen uniformly mixed, we expect
$0.1\times2.5=0.25\msun$ of oxygen to have been deposited on the 
companion, much more than the total mass it could have captured from
a spherically symmetry supernova. [Si/O] is 0.09 by mass in the Sun, and
[S/O] is 0.05, so since the over-abundances of all three elements are similar
we expect those ratios to hold here, giving about 0.02$\msun$ of captured Si
and 0.01$\msun$ of captured S. We therefore need a layer of stellar ejecta
to have been captured which has twice as much Si as S, at the same
time as having about 10 times more O. From fig.~\ref{fig1i}, we see
that this occurs nowhere in a normal
supernova, but does happen in the hypernova model of Nomoto et~al.\ (2000)
at mass cuts of 6$\msun$ or more. This agrees very nicely with the notion
that a hypernova took place in this system, and that the inner 7$\msun$ or
so went into a black hole.

What remains is to explain how the companion acquired ten times more mass
than the spherical supernova model allows, and once again we believe
that the answer is given in recent hypernova calculations (MacFadyen and
Woosley 1999, Wheeler et~al.\ 2000): hypernovae are powered by jet flows,
which means they are very asymmetric, with mass outflow
along the poles being much faster and more energetic than along the
equator. The disk provides a source for easily captured material
in two ways: First, it concentrates mass in the equatorial plane, which will
later be ejected mostly in that plane. Second, the velocity acquired by the
ejecta is of the order of the propagation speed of the shock through it. This
propagation speed is proportional to $\sqrt{P_2/\rho_1}$, where $P_2$ is
the pressure behind the shock and $\rho_1$ the density ahead of it. The driving
pressure will be similar in all directions (or larger, due to the jet injection,
in the polar regions), whereas the disk density is much higher than the polar
density. Hence, the equatorial ejecta will be considerably slower
than even normal supernova ejecta, greatly increasing the possibility of their
capture by the companion. Other significant effects of the disk/jet geometry
are (1) that the companion is shielded from ablation of its outer layers by fast
ejecta, which is thought to occur in spherical supernovae with companion
stars (Marietta, Burrows \& Fryxell 2000) and (2) that there is no iron
enrichment of the companion, because the iron ---originating closest to the
center--- is either all captured by the black hole or ejected mainly in the
jet, thus not getting near the companion (Wheeler et~al.\ 2000; note that 
indeed no overabundance of Fe is seen in the companion of GRO\,J1655$-$40).

For the companion to capture the required 0.2--0.3$\msun$ of ejecta it is
sufficient that the ejecta be slow enough to become gravitationally bound
to it. However, the material may not stay on: when the companion
has so much mass added on a dynamical time scale it will be pushed out of 
thermal equilibrium, and respond by expanding, as do main-sequence stars 
that accrete mass more gradually on a time scale faster than their thermal
time scale (e.g., Kippenhahn \& Meyer-Hofmeister 1977)\nocite{Kippenhahn77}.
During this expansion, which happens on a time scale much longer than the
explosion, the star may expand beyond its Roche lobe and transfer some of its
mass to the newly formed black hole. However, because the dense ejecta mix
into the envelope on a time scale between dynamical and thermal, i.e., faster
than the expansion time, this back transfer will not result in the bulk of the
ejecta being fed back, though probably the material lost is still richer in
heavy elements than the companion is now. Since the outer layers of the star
are not very dense, and the mass transfer is not unstable because the black
hole is much more massive than the companion, the total amount of mass 
transferred back is probably not dramatic. However, the expansion does 
imply that the
pre-explosion mass of the companion was somewhat higher than its present
mass, and that the amount of ejecta that needs to be captured in order to
explain the abundances observed today is also somewhat higher than the present
mass of heavy elements in the companion.

A further piece of evidence that may link Nova Sco 1994 to our GRB/hypernova
scenario are the
indications that the black hole in this binary
is spinning rapidly.  Zhang, Cui, \& Chen (1997)\nocite{Zhang97} argue from
the strength of the ultra-soft X--ray component that the black hole
is spinning near the maximum rate for a Kerr black hole.  However,
studies by Sobczak et al. (1999)\nocite{Sobczak99} show that it must be
spinning with less than 70\% maximum.  Gruzinov (1999)\nocite{Gruzinov99}
finds the inferred black hole spin to be about 60\% of maximal from
the 300 Hz QPO.  Our estimates of the last section indicate that enough
rotational energy will be left in the black hole so that it will still
be rapidly spinning.

We have already mentioned the unusually high space  velocity of
$-150 \pm 19$ km s$^{-1}$. Its origin was first discussed by 
Brandt et~al.\ (1995), who concluded that significant mass must have been
loss in the formation of the black hole in order to explain this high
space velocity:
it is not likely
to acquire a substantial velocity in its own original frame of
reference, partly because of the large mass of the black hole. But
the mass lost in the supernova explosion
 is ejected from a moving object and thus
carries net momentum. Therefore, momentum conservation demands that
the center of mass of the binary acquire a velocity;
this is the Blaauw--Boersma kick (Blaauw 1961, Boersma
1961)\nocite{Blaauw61,Boersma61}. 
Note that the F-star companion mass is the largest among the black-hole
transient sources, so the center of mass is furthest from the black hole
and one would expect the greatest kick.
Nelemans et al. (1999) estimate
the mass loss in this kick to be $5-10\msun$.

In view of the above, we consider it well established that Nova Sco
1994 is the relic of a hypernova. We believe it highly likely that the
other black-hole transient X-ray sources are also hypernova remnants. 
We believe it
likely that the hypernova explosion was accompanied by a GRB if, as in
GRB\,980326, the energy was delivered in a few seconds. It is not clear
 what will happen if the magnetic fields are so low that the power is
delivered only over a much longer time. There could then still be intense
power input for a few seconds due to neutrino annihilation deposition
 near the black hole (Janka et~al.\ 1999), but that may not be enough
 for the jet to pierce through the He star and cause a proper GRB
(MacFadyen and Woosley 1999). At
this point, we recall that the GRB associated with SN1998bw was very
sub-luminous, $10^5$ times lower than most other GRB. While it has been
suggested that this is due to us seeing the jet sideways, it is
in our view more likely that the event was more or less spherical
(Kulkarni et~al.\ 1998)
and we see a truly lower-power event. A good candidate would be the
original suggestion by Colgate (1968, 1974) of supernova shock break-out
producing some gamma rays. Indications are that the expansion in SN1998bw
was mildly relativistic (Kulkarni et~al.\ 1998) or just 
sub-relativistic (Waxman and Loeb 1999). In either
case, what we may have witnessed is a natural intermediate event in our
scenario: we posit that there is a  continuum of events varying from
normal supernovae, delivering 1\,foe more or less spherically in ten
seconds, to extreme hypernovae/GRB that deliver 100 foes in a highly
directed beam. In the middle, there will be cases where the beam cannot
pierce through the star, but the total energy delivered is well above a
supernova, with as net result a hypernova accompanied by a very weak GRB.

\subsection{Numbers}

Nearly all observed black hole transient X-ray sources are within
5\,kpc of the Sun. Extrapolating to the entire Galaxy, a total of
8,800 black-hole transients with main-sequence K companions
has been suggested (Brown, Lee, \& Bethe 1999).

The lifetime of a K star in a black hole transient X-ray source is
estimated to be $\sim 10^{10}$\,yr (Van Paradijs 1996) but we
shall employ $10^9$\,yr for the average of the K-stars and the more
massive stars, chiefly those in the ``silent partners". In this
case the birth rate of the observed transient sources would be
   \be
   \lambda_K=10^4/10^9=10^{-5} {\rm per \ galaxy\ yr^{-1}}.
   \ee

We see no reason why low-mass companions should be preferred, so
we assume that the formation rate of binaries should be
independent of the ratio
   \be
   q=M_{B,i}/M_{A,i}.
   \ee
In other discussions of binaries, e.g., in Portegies Zwart \&
Yungelson (1998), it has often been assumed that the distribution
is uniform in $q$. This is plausible but there is no proof. Since
all primary masses $M_A$ are in a narrow interval, 20 to
$35\msun$, this means that $M_B$ is uniformly distributed between
zero and some average $M_A$, let us say $25\msun$. Then the total
rate of creation of binaries of our type is
  \be
  \lambda=\frac{25}{0.7}\lambda_K=3\times 10^{-4} {\,\rm galaxy^{-1}\ yr^{-1}}.
  \label{eq7.17}
  \ee
This is close to the rate of mergers of low mass black holes with neutron
stars which Bethe \& Brown (1998) have estimated to be
  \be
  \lambda_m \simeq 2\times 10^{-4}  {\,\rm galaxy^{-1}\ yr^{-1}}.
  \label{eq7.18}
  \ee
These mergers have been associated speculatively with short GRBs, while
formation of our binaries is supposed to lead to ``long" GRBs 
(Fryer, Woosley, \& Hartmann 1999). We
conclude that the two types of GRB should be equally frequent,
which is not inconsistent with observations. In absolute number both of our
estimates eqs.~(\ref{eq7.17}) and (\ref{eq7.18}) are
substantially larger than the observed rate of
$10^{-7}$ galaxy$^{-1}$ yr$^{-1}$ (Wijers et al. 1998); this is natural,
since substantial beaming is expected in GRBs produced by the
Blandford-Znajek mechanism.
Although we feel our mechanism to be fairly general, it may be that the
magnetic field required to deliver the BZ energy within a suitable time
occurs in only a fraction of the He cores.

\section{Discussion and Conclusion}
\label{conclu}

Our work here has been based on the Blandford-Znajek mechanism of
extracting rotational energies of black holes spun up by accreting
matter from a helium star. We present it using the simple circuitry of {\it
``The Membrane Paradigm"} (Thorne et~al.\ 1986)\nocite{Thorne86}. Energy
delivered into the loading region up the rotational axis of the black
hole is used to power a GRB. The energy delivered into the accretion disk
powers a SN Ib explosion.

We also discussed black-hole transient sources, high-mass black holes
with low-mass companions, as possible relics for both GRBs
and Type Ib supernova explosions, since there are indications that
they underwent mass loss in a supernova explosion. In Nova Sco
1994 there is evidence from the atmosphere of the companion star
that a very powerful supernova explosion (`hypernova') occurred.

We estimate the progenitors of transient sources to be formed at a
rate of 300 GEM (Galactic Events per Megayear).  Since this is
much greater than the observed rate of GRBs, there must be strong
collimation and possible selection of high magnetic fields in order to
explain the discrepancy.

We believe that there are strong reasons that a GRB must be associated
with a black hole, at least those of duration several seconds or more
discussed here. Firstly, neutrinos can deliver energy from a stellar
collapse for at most a few seconds, and sufficient power for at most
a second or two. Our quantitative estimates show that the rotating black
hole can easily supply the energy as it is braked, provided the ambient
magnetic field is sufficiently strong. The black hole also solves
the baryon pollution problem: we need the ejecta that give rise to the
GRB to be accelerated to a Lorentz factor of 100 or more, whereas the
natural scale for any particle near a black hole is less than its mass.
Consequently, we have a distillation problem of taking all the energy
released and putting it into a small fraction of the total mass.
The use of a Poynting flux from a black hole in a magnetic field
(Blandford \& Znajek 1977) does not require the presence of much mass,
and uses the rotation energy of the black hole, so it provides
naturally clean power.

Of course, nature is extremely inventive, and we do not claim that 
all GRBs will fit into the framework outlined here. We would not
expect to see all of the highly beamed jets following from the BZ mechanism
head on, the jets may encounter some remaining hydrogen envelope in some
cases, jets from lower magnetic fields than we have considered here may
be much weaker and delivered over longer times, etc., so we speculate that
a continuum of phenomena may exist between normal supernovae and 
extreme hypernovae/GRBs. This is why we
call our effort ``A Theory of Gamma Ray Bursts" and hope that it will
be a preliminary attempt towards systematizing the main features of
the energetic bursts.

\acknowledgements

We would like to thank Stan Woosley for much useful information.
Several conversations with Roger Blandford made it possible for
us to greatly improve our paper, as did valuable comments from
Norbert Langer.
This work is partially supported by the U.S. Department of Energy
Grant No. DE-FG02-88ER40388. HKL is supported  also in part by 
KOSEF Grant No.\
1999-2-112-003-5 and by the BK21 program of the Korean Ministry of
Education.

\appendix

\section{Estimates of $\epsilon_\Omega=\Omega_{disk}/\Omega_H$}

We collect here useful formulas needed to calculate
$\epsilon_\Omega=\Omega_{disk}/\Omega_H$.
First of all
   \be
   \Omega_H &=&\frac{\tilde a}{1+\sqrt{1-\tilde a^2}}
               \left( \frac{c^3}{2MG}\right)
   \nonumber\\
   &=&\frac{\sqrt 2 \ \tilde a}{1+\sqrt{1-\tilde a^2}}
      \left(\frac{R}{R_{Sch}}\right)^{3/2}\Omega_K
   \\
   \Omega_{disk} &=& \omega \ \Omega_K \left[\ 1+\tilde a\frac{GM}{c^2}
                       \sqrt{\frac{GM}{c^2R^3}}\ \right]^{-1}
   \nonumber\\
   &=& \omega \ \Omega_K \left[\ 1+\tilde a\left(
              \frac{R_{Sch}}{2R}\right)^{3/2}\  \right]^{-1}
   \ee
where $\Omega_K \equiv \sqrt{GM/R^3}$ and $\omega$ is
dimensionless parameter ($0<\omega <1$). Thus
   \be
   \frac{\Omega_{disk}}{\Omega_H}
   &=& \omega \
   \frac{1+\sqrt{1-\tilde a^2}}{\sqrt 2 \ \tilde a}
      \left(\frac{R_{Sch}}{R}\right)^{3/2}
    \left[\ 1+\tilde a\left(
              \frac{R_{Sch}}{2R}\right)^{3/2}\ \right]^{-1}.
   \ee
The numerical estimates are summarized in Table~\ref{appBtab1}
for various $\omega$ and radii.

\section{Spin-up of Black Holes by Accretion}

The specific angular momentum and energy of test particles in Keplerian
circular motion, with rest mass $\delta m$, are
  \be
  \tilde E &\equiv& \frac{E}{\delta m} = c^2 \left[
         \frac{r^2- R_{\Sch} r + a\sqrt{R_{\Sch}r/2}}
              {r(r^2-\frac32 R_{\Sch} r+ a\sqrt{2 R_{\Sch} r})^{1/2}} \right]
  \nonumber\\
  \tilde l &\equiv& \frac{l}{\delta m} = c
              \sqrt{\frac{R_{\Sch}r}{2}}
              \left[ \frac{(r^2- a \sqrt{2 R_{\Sch} r} + a^2)}
              {r(r^2-\frac32 R_{\Sch} r+ a\sqrt{2 R_{\Sch} r})^{1/2}} \right]
  \ee
where $R_{\Sch}=2GM/c^2$ and BH spin $a=J/Mc=\tilde a (GM/c^2)$.
The accretion of $\delta m$ changes the BH's total mass and angular
momentum by $\Delta M=\tilde E\delta m$ and $\Delta J=\tilde l \delta m$.
The radii of marginally bound ($r_{mb}$) and stable ($r_{ms})$ orbits
are given as
  \be
  r_{mb} &=& R_{\Sch}-a+\sqrt{R_{\Sch} (R_{\Sch}-2a)} \nonumber\\
  r_{ms} &=& \frac{R_{\Sch}}{2}\left(3+Z_2-\left[(3-Z_1)
   (3+Z_1+2 Z_2)\right]^{1/2}\right) \nonumber\\
  Z_1 &=& 1+\left(1-\frac{4 a^2}{R_{\Sch}^2}\right)^{1/3}
          \left[\left(1+\frac{2a}{R_{\Sch}}\right)^{1/3}
                +\left(1-\frac{2a}{R_{\Sch}}\right)^{1/3}\right] \nonumber\\
  Z_2 &=& \left(3\frac{4 a^2}{R_{\Sch}^2}+Z_1^2\right)^{1/2}.
  \ee
The numerical values of the specific angular momentum and energy of test
particles are summarized in Table \ref{Apptab1} and
Fig.\ref{Appfig1}.
In Fig.\ref{Appfig2},
we test how much mass we need in order to spin up
the non-rotating black hole up to given $\tilde a$.
Note that the last stable orbit is almost Keplerian even with the
accretion disk, and we assume 100\% efficiency of angular momentum transfer
from the last stable Keplerian orbit to BH.
In order to spin-up the BH up to $\tilde a=0.9$,
we need $\sim 68\%$ ($52\%$) of original non-rotating
BH mass in case of $r_{lso}=r_{ms}$ ($r_{mb}$).
For a very rapidly rotating BH with $\tilde a=0.99$, we need $122\%$
and 82\%, respectively.
For $r_{lso}=r_{ms}$, there is an upper limit, $\tilde a=0.998$, which can
be obtained by accretion (Thorne 1974). In the limit where $r_{lso}=r_{mb}$,
however, spin-up beyond this limit is possible because the photons can be
captured inside thick accretion disk, finally into BH
(Abramowicz et al. 1988).


%
%


\begin{table}
\begin{center}
\begin{tabular}{ccccc}
\hline
Spectral Type    & $M_B [\msun]$ &   $a [\rsun]$  &  $R_L [\rsun]$ & $j_{B,orb}$ [$10^{19}$ cm$^2$ s$^{-1}$] \\
\hline
5 K-type         &  $0.4-0.9$    & $4.5\pm 0.8$   & $0.9\pm 0.2$   & $1.5\pm 0.6$ \\
1 K-type         &    0.8        &   34           &  6.1           &  5.7    \\
1 M-type         &    0.5        &   3.1          &  0.6           &  1.5    \\
F (Nova Scorpii) &    2.2        &   16           &  4.7           &  1.9    \\
A2 (1543-47)     &    2.0        &   9.0          &  2.6           &  1.4    \\
\hline
\end{tabular}
\end{center}
\caption{Properties of transient X-ray sources}
\label{tabhans}
\end{table}

\begin{table}
$$
\begin{array}{|cc|cccc|}
\hline
      &      & \multicolumn{4}{c|}{\Omega_{disk}/\Omega_H}\\
\omega  & \tilde a & r=r_{mb}(\tilde a) & r=r_{ms}(\tilde a) &
                     r=2 R_{Sch} & r= 3 R_{Sch} \\
\hline
1.0   & 0.80 & 1.00 & 0.69 & 0.45 & 0.26 \\
0.9   & 0.72 & 0.99 & 0.63 & 0.49 & 0.27 \\
0.8   & 0.64 & 0.93 & 0.58 & 0.51 & 0.29 \\
0.7   & 0.56 & 0.89 & 0.54 & 0.53 & 0.30 \\
0.6   & 0.48 & 0.84 & 0.50 & 0.55 & 0.31 \\
0.5   & 0.40 & 0.80 & 0.46 & 0.57 & 0.32 \\
\hline
\end{array}
$$
\caption{Estimates of $\epsilon_\Omega=\Omega_{disk}/\Omega_{H}$ as a function
of spin parameter and radius, where $r_{mb}$ is the marginally bound
radius and $r=r_{ms}$ the marginally stable radius.}
\label{appBtab1}
\end{table}

\newpage
\renewcommand{\thetable}{\arabic{table}}
\renewcommand{\thefigure}{\arabic{figure}}

\begin{table}
$$
\begin{array}{ccccc}
\hline
          & a [GM/c^2] &\tilde l [GM/c] & r [R_{\Sch}] & \tilde\epsilon [c^2]\\
\hline r_{ms} & 0 & 2\sqrt{3}\approx 3.46 &   3    & \sqrt{8/9}\approx 0.943\\
              & 1 & 2/\sqrt{3}\approx 1.15 &  1    & \sqrt{1/3}\approx 0.577\\
\hline r_{mb} & 0 &               4        &  2    &  1 \\
              & 1 &               2        &  1    &  1 \\
\hline
\end{array}
$$ \caption{Properties of Schwarzschild \& Kerr BH. a)
$r_{lso}=r_{ms}$ case : 6\% (42\%) of energy
can be released during the spiral-in for Schwarzschild
(maximally-rotating Kerr) BHs. b) $r_{lso}=r_{mb}$ case : The released
energy during the spiral-in is almost zero. }
\label{Apptab1}
\end{table}

%
%

\begin{figure}
\centerline{\epsfig{file=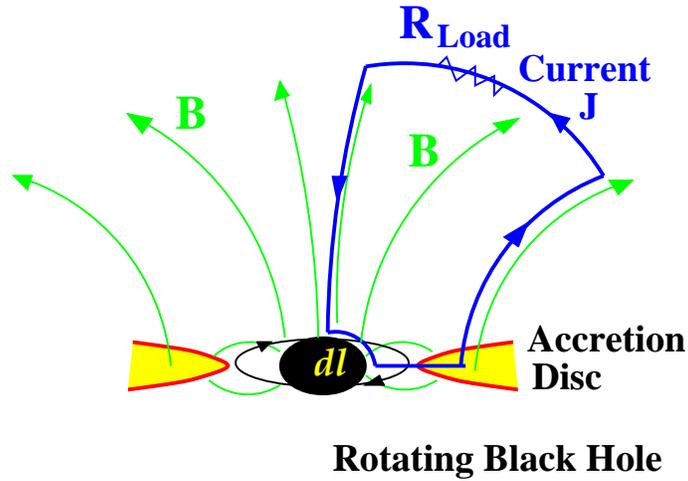,height=2.5in}} \caption{The
black hole in rotation about the accretion disk. A circuit,
in rigid rotation with the black hole is shown. This
circuit cuts the field lines from the disk as the black hole
rotates, and by Faraday's law, produces an electromotive force.
This force drives a current. More detailed discussion is given in
the text.
\label{bzcirc}}
\end{figure}

\begin{figure}
\centerline{\epsfig{file=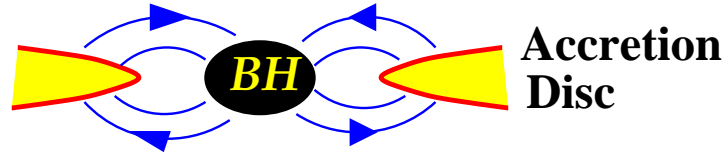,height=1in}}
\caption{Magnetic field lines, anchored in the disk, which thread
the black hole, coupling the disk rotation to that of the black hole.}
\label{mbreak}
\end{figure}

\begin{figure}
\centerline{\epsfig{file=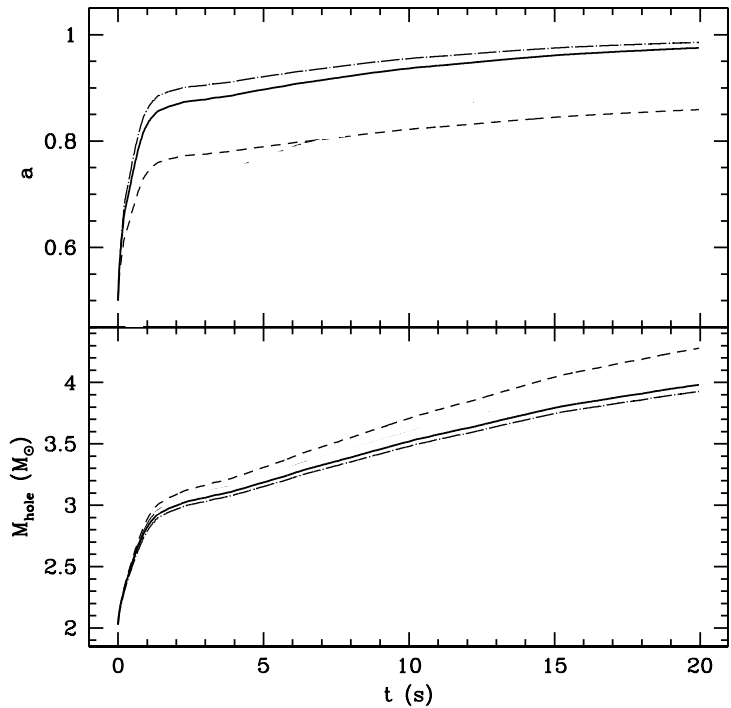,height=4.0in}}
\caption{Time evolution of BH mass and angular momentum taken
from Fig. 19 of MacFadyen \& Woosley 1999.
The upper panel shows the increase in the Kerr parameter for various
models for the disk interior to the inner boundary at 50 km.
``Thin" (dash-dot), neutrino-dominated (thick solid) and advection
dominated (short dash) models are shown for initial Kerr parameter
$\tilde a_{init}=0.5$.
The lower panel shows the growth of the gravitational mass of the
black hole. The short-dashed line shows the growth in baryonic mass of
the black hole since for a pure advective model no energy escapes the
inner disk.
}
\label{woosley}
\end{figure}

\begin{figure}
\epsfig{file=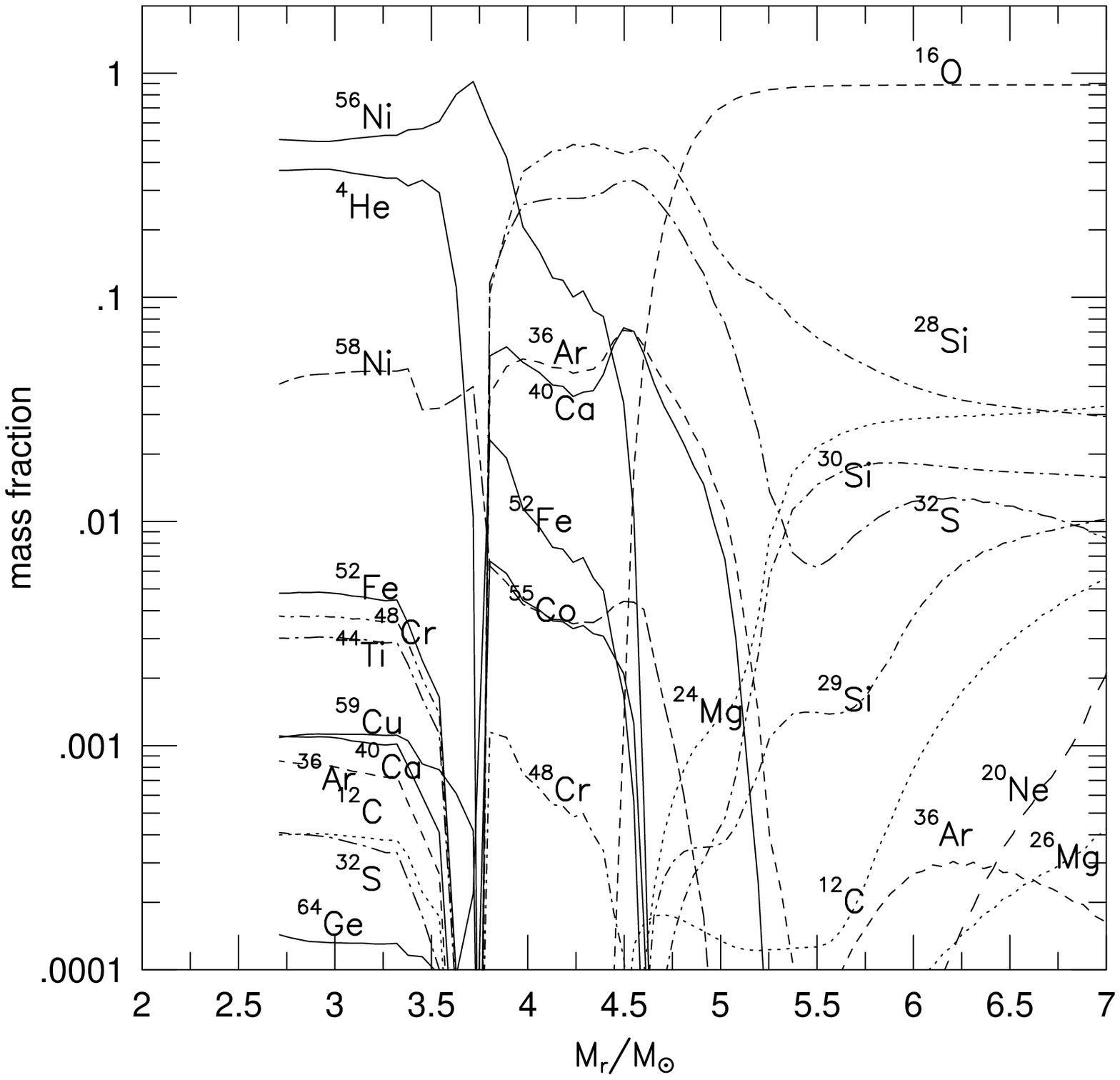,width=0.48\textwidth}\hfill\epsfig{file=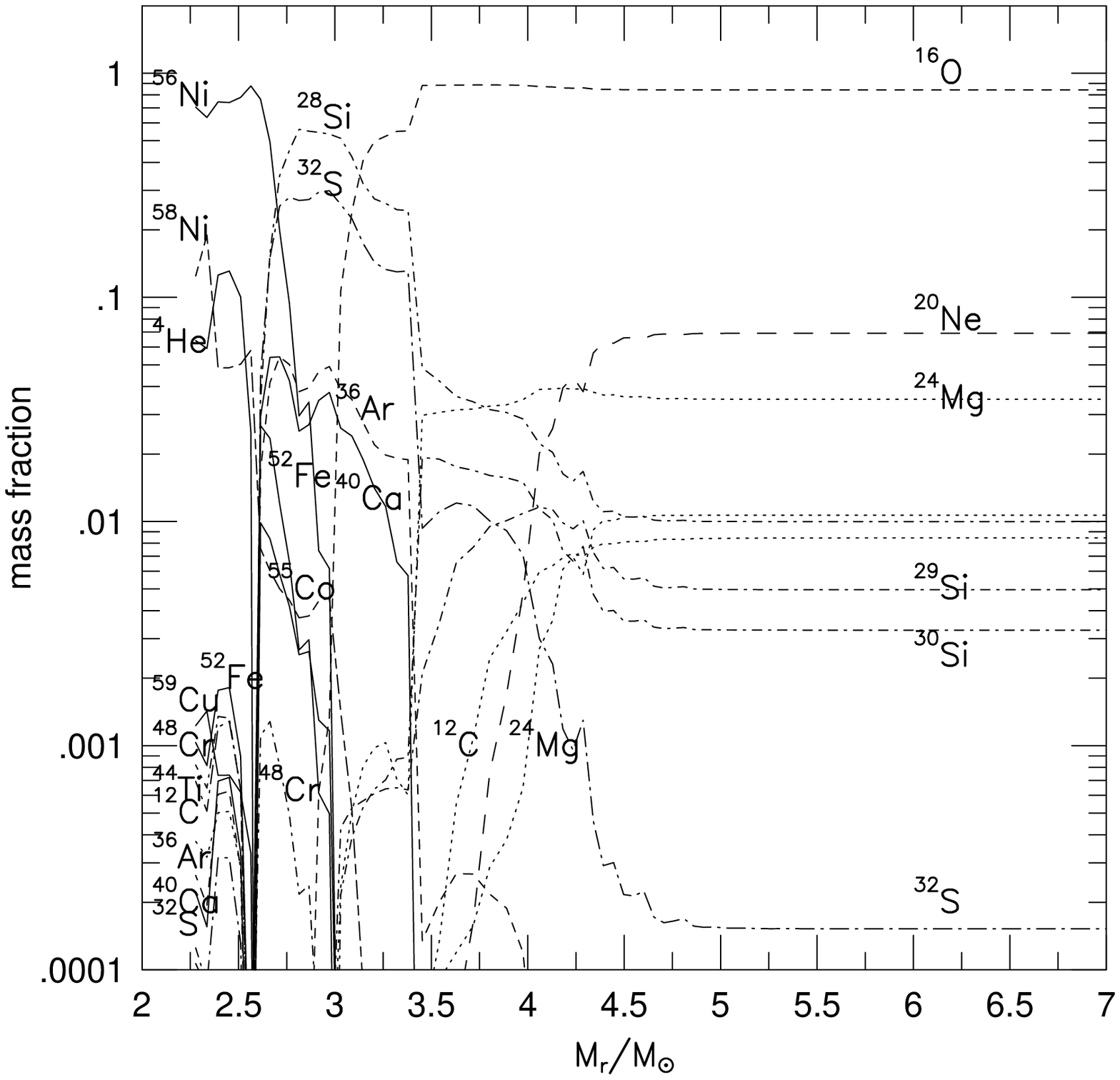,width=0.48\textwidth}
\caption{
The isotopic composition of ejecta of the hypernova 
($E_{\rm K}=3\times10^{52}$\,erg; left) and the normal supernova 
($E_{\rm K}=1\times10^{51}$\,erg; right)
for a $16\msun$ He star, from Nomoto et~al.\ (1999). Note the much higher 
sulphur abundance in the hypernova.
}\label{fig1i}
\end{figure}

\begin{figure}
\begin{center}
\epsfig{file=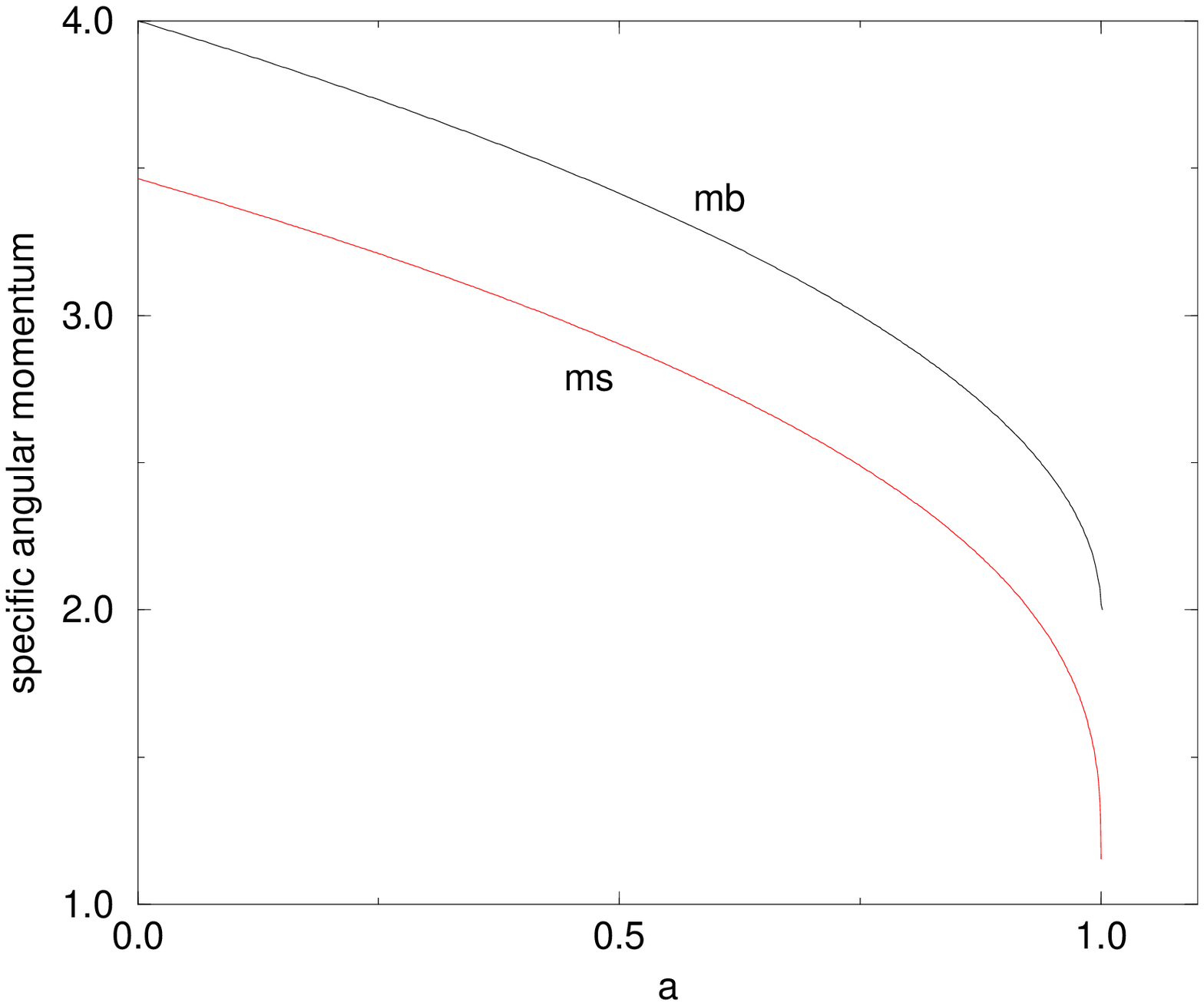,width=4.2in}
\epsfig{file=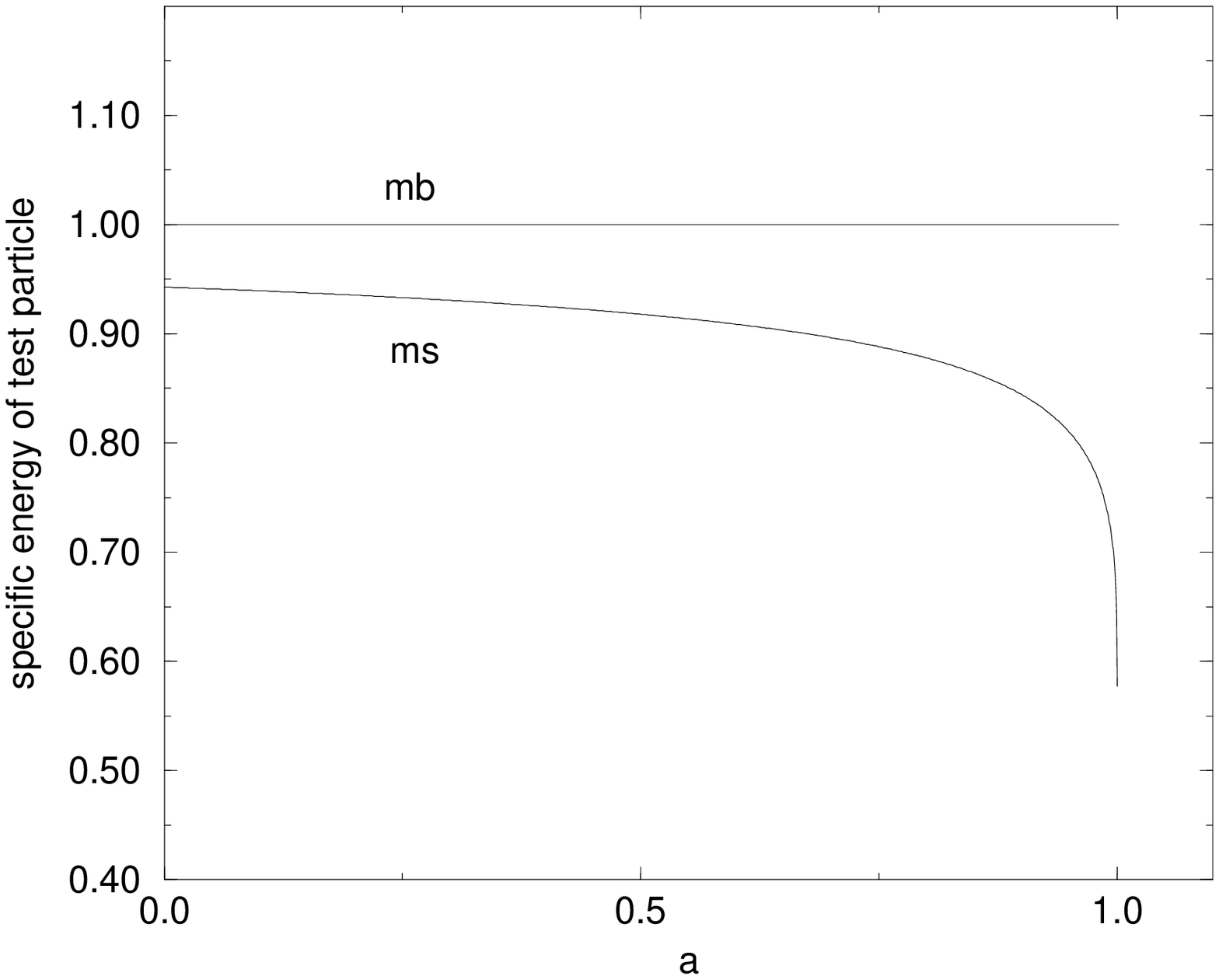,width=4.2in}
\end{center}
\caption{Specific angular momentum and energy
of test particle in units of [$GM/c$] and [$c^2$].
BH spin $a$ is given in unit of [$GM/c^2$].
For the limiting values at $a=0$ and $GM/c^2$, refer Table \ref{Apptab1}.
  }
\label{Appfig1}
\end{figure}

\begin{figure}
\begin{center}
\epsfig{file=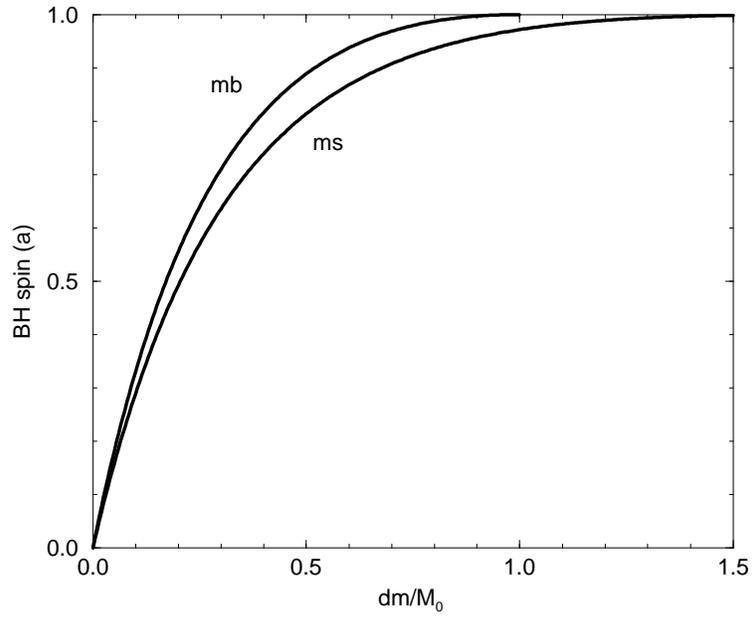,width=4.5in}
\end{center}
\caption{Spinning up of Black Holes. The
BH spin $a$ is given in units of [$GM/c^2$] and $\delta m$ is
the total rest mass of the accreted material.
}
\label{Appfig2}
\end{figure}

\end{document}